\documentclass[aps,pre,twocolumn,groupedaddress]{revtex4}
\usepackage{graphicx}
\begin{document}

\title{Melting and unzipping of DNA}

\author{Y. Kafri and D. Mukamel}
\affiliation{Department of Physics of Complex Systems, The
Weizmann Institute of Science, Rehovot 76100, Israel}

\author{L. Peliti}
\affiliation{Dipartimento di Scienze Fisiche and Unit\`a INFM,
Universit\`a ``Federico II'', Complesso Monte S. Angelo, I--80126
Napoli, Italy}

\date{Aug 21, 2001}

\begin{abstract}
Experimental studies of the thermal denaturation of DNA yield a
strong indication that the transition is first order. This
transition has been theoretically studied since the early sixties,
mostly within an approach in which the microscopic configurations
of a DNA molecule are given by an alternating sequence of
non-interacting bound segments and denaturated loops. Studies of
these models neglect the repulsive, self-avoiding, interaction
between different loops and segments and have invariably yielded
continuous denaturation transitions. In this study we exploit
recent results on scaling properties of polymer networks of
arbitrary topology in order to take into account the
excluded-volume interaction between denaturated loops and the rest
of the chain. We thus obtain a first-order phase transition in
$d=2$ dimensions and above, in agreement with experiments. We also
consider within our approach the unzipping transition, which takes
place when the two DNA strands are pulled apart by an external
force acting on one end. We find that the unzipping transition is
also first order. Although the denaturation and unzipping
transitions are thermodynamically first order, they do exhibit
critical fluctuations in some of their properties. For instance,
the loop size distribution decays algebraically at the transition
and the length of the denaturated end segment diverges as the
transition is approached. We evaluate these critical properties
within our approach.
\end{abstract}


\maketitle


\section{Introduction}
The unbinding phase transition of the two complementary strands of
the DNA molecule has been a subject of continual interest for over
four decades \cite{WB,PS1,Lifson,PB,CH,CM,TDP}. In thermal
denaturation this transition takes place when temperature is
increased. In a typical experiment, a sample containing DNA
molecules of specific length and sequence is prepared. The
fraction of attached bound base pairs, $\theta$, is measured
through light absorption at a wavelength of 260 nm. At low
temperatures all base pairs are attached to each other while at
high temperature they are all unbound. Thus, $\theta$ decreases
from one to zero as the temperature is increased. For
heterogeneous DNA molecules, containing both AT and GC base pairs,
$\theta$ does not decrease smoothly with temperature, but rather
exhibits a multistep behavior. It consists of plateaus of various
sizes separated by sharp jumps. This behavior is related to the
fact that the GC bonds are stronger than AT ones. Thus, long
domains with higher concentration of AT bonds will denaturate at
lower temperatures. The resulting stepped structure of $\theta$ is
therefore characteristic of the particular DNA sequence. It thus
yields statistical information on the sequence of the molecule
under study. The denaturation process has been verified through
electron microscopy where denaturate loops and bound segments have
been observed directly \cite{PLBL}. The sharpness of the jumps
indicates that the unbinding transition is {\it first order}.

More recently, the introduction of new techniques such as optical
tweezers and atomic force microscopy \cite{tweezers,AFM} has
allowed the manipulation of single biological molecules. This made
it possible to study a wider variety of physical properties of the
DNA molecule. For example optical tweezers have been used to apply
a force and pull apart the two strands at one end of the molecule.
It is found that a phase transition takes place at a critical
force where the molecule is unzipped and the two strands are
separated \cite{Brockelmann}.

Thermal denaturation has been studied theoretically since the
early sixties. The early models, which we refer to as
Poland-Scheraga (PS) type models \cite{PS1,Lifson}, consider the
molecule as being composed of an alternating sequence of bound and
denaturated segments. A bound segments is energetically favored
over an unbound segment, while a denaturated segment (loop) is
entropically favored over a bound one. Within this approach the
interaction between the different segments of the molecule has not
been taken into account. This assumption simplifies the analysis
considerably. The order of the transition is found to be
determined by a parameter $c$ which characterizes the statistical
weight of a loop. The number of configurations of a loop of length
$\ell$ behaves as $s^\ell/\ell^c$ for large $\ell$. Here $s$ is a
non-universal constant. It has been shown \cite{PS2} that the
phase transition is first order if $c>2$ and second order if $1< c
\leq 2$ while for $c<1$ no transition takes place and the strands
are always bound. Using random-walk configurations to model a loop
one finds that $c=d/2$ in $d$ dimensions. The transition is thus
predicted to be continuous in $d=2$ and $d=3$ dimensions
\cite{PS2}. This result is at variance with experimental
observations. The model was later generalized to take into account
the self-avoiding interactions {\it within} each loop
\cite{Fisher,COM}. It is found that the loop entropy takes the
same general form as before. However, the exponent $c$ now takes
the value $d \nu$, where $\nu$ the correlation length exponent of
a self-avoiding random walk. Inserting the known values for $\nu$
one finds that although $c$ is larger than that of a random-walk
model, it is still smaller than two both in $d=2$ and $d=3$
dimensions, yielding a continuous transition. It was suggested
\cite{Fisher} that self-avoiding interactions {\it between} the
various parts of the chain (and not just within loops) would
further sharpen the transition possibly making it first order.
However, theoretical tools for carrying out this analysis have not
been available at the time. More recently, excluded volume
interactions have been fully taken into account in a numerical
study of finite chains \cite{CCG}. These simulations strongly
suggest that the transition is indeed first order.

In a recent study \cite{KMP}, we have extended the PS model to
take into account self-avoiding interactions both within a loop
and between a loop and the rest of the chain. To carry out the
analysis of this model one has to enumerate the configurations of
a loop embedded in a chain with self-avoiding interactions. This
has been done by taking advantage of recent results obtained by
Duplantier et~al.~\cite{Dup,SFLD} for the number of configuration
of a general polymer network. It is found that the statistical
weight of a loop embedded in a chain has the same general form as
before, namely $s^\ell/\ell^c$. However, the parameter $c$ is now
modified and becomes larger than two in $d \geq 2$ dimensions. In
particular one finds $c \simeq 2.115$ in $d=3$ and $c = 2+13/32$
in $d=2$. Thus, self-avoiding interactions make the transition
first order in two dimensions and above. Recently a different
model in which excluded volume interactions were partially taken
into account has been found to yield a first order transition
\cite{GMO}. In this model excluded volume interactions between the
two strands of the chain were explicitly considered, but those
within each strand were neglected.

In this work we present a detailed account of the results obtained
when the interactions between various segments along the chain are
taken into account using the scaling results of Duplantier. We
show that the denaturation transition is first order. However, the
transition is found to be accompanied by critical fluctuations in
some the chain's properties. For example, the loop size
distribution is found to decay algebraically at the transition.
Indeed, the probability distribution for loops of length $\ell$,
$P(\ell)$, behaves as $P(\ell) \sim 1/\ell^c$ at the transition.
This behavior was recently confirmed in numerical simulations of
the model where the excluded volume interactions have been taken
into account fully. The value of the measured exponent $c \sim
2.10 \pm 0.02$ agrees well with our predictions \cite{COS}. We
also find that when the boundary conditions are such that the
chain is open at one end the length $\xi$ of the end segment
diverges as $1/|T-T_{\rm M}|$ when the melting temperature $T_{\rm
M}$ is approached.

We have extended the model to consider the unzipping transition
which takes place when a force of magnitude $f$ is applied to
separate the two strands. We find that this transition is first
order. We also find that the end segment length diverges as $\xi
\sim 1/|f-f_{\rm U}|$ where $f_{\rm U}$ is the unzipping critical
force. This behavior has been previously found in models where
self avoiding interactions were not taken into account
\cite{BHA,LN,LN2,MIM,CMM}. In calculating the critical force near
the melting transition we find that $f_{\rm U} \sim |T-T_{\rm
M}|^\nu$.

The paper is organized as follows. In Section II we consider the
thermal denaturation transition in detail. In Section IIA we
review the analysis of Poland and Scheraga and that of Fisher
where self-avoiding interaction within a loop is taken into
account. In Section IIB we analyze the model with full account of
self-avoiding interactions. The length distribution of the end
segment is considered in section IIC and a summary and overview of
other approaches to thermal denaturation of DNA is given in
Section IID. In Section III the unzipping transition is studied. A
brief summary is given in section IV.

\section{Thermal Denaturation of DNA}

\subsection{The model and basic analysis}

The model considers two strands, each composed of monomers. Each
monomer represents one persistence length of a single unbound
strand. Typically this is about $\sim 40$~\AA\ \cite{RMMRSU}, or
roughly 8 bases. We set the boundary conditions such that the
monomers at one end of the molecule are bound. Such a boundary
condition is necessary for a bound state between the two strands
to exist. All other monomers on the chain can be either unbound or
bound to a specific matching monomer on the second chain. The
interactions between a monomer and other monomers on the second
strand or on the same strand are ignored. The binding energy
$E_0<0$ between matching monomers is taken to be the same for all
monomer pairs.

\begin{figure}
\includegraphics[width=8cm]{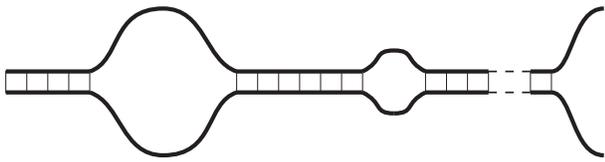} \caption{Schematic
representation of a microscopic configuration of the DNA molecule.
\label{fig:model}}
\end{figure}

A typical DNA configuration is shown in Fig.~\ref{fig:model}. It
is made of an alternating sequence of bound segments and
denaturated loops. The configuration ends with two denaturated
strands. For simplicity, the configurational entropy of a bound
segment associates with its embedding in ambient space is
neglected. It is easy to verify that this assumption does not
affect the nature of the denaturation transition obtained within
this model. The statistical weight of a bound sequence of length
$\ell$ is then given by $ w ^\ell = \exp(- \ell E_0/T)$, where $T$
is the temperature and the Boltzmann constant $k_{\rm B}$ is set
to 1. Thus, $w$ is a decreasing function of the temperature. On
the other hand, a denaturated loop does not carry an energy and
its statistical weight is derived from its degeneracy. In this
model it is assumed that the loop is fully flexible, and thus it
is described by a random walk which returns to the origin after $2
\ell$ steps. Considering all possible such walks the statistical
weight for large $\ell$ has the form $ \Omega(2 \ell ) = A s^\ell
/ \ell^c$, where $s$ is a non-universal constant and the exponent
$c$ is determined by the properties of the loop configurations.
For simplicity, we set $A=1$. Finally, the statistical weight of
the end segment, which consists of two denaturated strands each of
length $\ell$, takes the form $\Lambda (2 \ell )= B s^\ell/
\ell^{\bar{c}}$ for large $\ell$, where $\bar{c}$ is in general
not equal to $c$. Again, for simplicity we set $B=1$. The values
of the exponents $c$ and $\bar{c}$ will be held arbitrary for the
moment. We shall later discuss them in detail.

Using the weights assigned to each segment of the chain the total
weight of any given configuration may be calculated. For example
the weight of a chain which consists of a bound segment of length
$\ell_1$, a denaturated loop of length $\ell_2$, a bound segment
of length $\ell_3$, and a pair of denaturated strands of length
$\ell_4$, is given by
\begin{equation}
w ^{\ell_1} \cdot  \Omega(2 \ell_2) \cdot  w ^{\ell_3} \cdot
\Lambda(2 \ell_4). \label{samconf}
\end{equation}
The statistical weight of a more general chain configuration made
of $p$ alternating bound segments and denaturated loops will have
the same form with a suitable number of factors of the form $
\Omega(2 \ell_i) \cdot w ^{\ell_{i+1}}$ before the end-segment
weight $\Lambda(2 \ell_p)$.

\begin{figure}
\includegraphics[width=8cm]{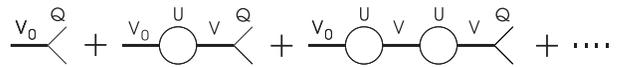} \caption{Graphic illustration of the expansion of the partition
function (\ref{partition}) in $U(z)V(z)$. Each segment of type
$U_0$,$U$,$V$ or $Q$ represents a sum over all possible lengths of
its type weighted properly with a fugacity. \label{fig:dnaseries}}
\end{figure}

The model is most easily studied within the grand canonical
ensemble where the total chain length $L$ is allowed to fluctuate.
The grand canonical partition function, ${\cal Z}$, is given by
\begin{equation}
{\cal Z}=
\sum_{L=0}^{\infty}Z(L)\,z^L=\frac{V_0(z)Q(z)}{1-U(z)V(z)},
\label{partition}
\end{equation}
where $Z(L)$ is the canonical partition function of a chain of
length $L$, $z$ is the fugacity, the functions $U(z)$, $V(z)$
and $Q(z)$ are defined by
\begin{mathletters}
\begin{eqnarray}
\label{UV} U(z)&=& \sum_{\ell=1}^{\infty} \Omega(2\ell) z^\ell
=\sum_{\ell=1}^{\infty}\frac{s^\ell}{\ell^c}\,z^\ell
=\Phi_c(zs), \label{Udef}\\
V(z)&=& \sum_{\ell=1}^{\infty} w ^\ell z^\ell, \label{Vdef} \\
Q(z)&=& 1+\sum_{\ell=1}^{\infty} \Lambda(2\ell) z^\ell
=1+\sum_{\ell=1}^{\infty}
\frac{s^\ell}{\ell^{\bar{c}}}\,z^\ell\nonumber\\
&=& 1+\Phi_{\bar{c}}(zs),
\end{eqnarray}
\end{mathletters}
\noindent and $V_0(z)=1+V(z)$. In this equation, $\Phi_c(z)$ is
the polylog function whose basic properties are summarized in
Appendix~\ref{polylog:sec}. Equation (\ref{partition}) can be
verified by expanding the partition function as a series in
$U(z)V(z)$. The factors $V_0(z)$ and $Q(z)$ properly account for
the boundaries. A graphical illustration of the series expansion
in $U(z)V(z)$ is given in Fig.~\ref{fig:dnaseries}. To set the
average chain length, $\langle L \rangle$, one has to choose a
fugacity such that
\begin{equation}\label{fugacity:eq}
\langle L \rangle=\partial \ln {\cal Z}/\partial \ln z.
\end{equation}
This implies that the thermodynamic limit $\langle L \rangle \to
\infty$ is obtained by letting $z$ approach the lowest fugacity
$z^*$ for which the partition function (\ref{partition}) diverges.
This can arise either from the divergence of the numerator or from
the vanishing of the denominator. The relevant situation, at low
temperature, is the second one, which corresponds to $z^*$
satisfying
\begin{equation}
U(z^*)V(z^*)=1. \label{pole}
\end{equation}
Since $V(z)= w  z /(1-  w  z)$ this reduces to
\begin{equation}
U(z^*)=1/( w  z^*)-1. \label{transtion}
\end{equation}
We shall see that above the transition, namely in the denaturated
phase, the numerator diverges. Moreover, when one considers the
problem of DNA unzipping by applying an external force on the
strands, a divergence arising from a boundary factor will play an
important role.

The fraction of bound monomer pairs $\theta$ is the experimentally
measured quantity and the order parameter of the transition. Its
temperature dependence in the thermodynamic limit $\langle L
\rangle \to \infty$ can be calculated from the behavior of
$z^*(w)$. The average number of bound pairs in a chain is given by
$\langle m \rangle=
\partial \ln {\cal Z}/\partial \ln  w $, so that
\begin{equation}
\theta= \lim_{L \to \infty} \frac{\langle m \rangle}{\langle L
\rangle} =\frac{
\partial \ln z^* }{\partial \ln  w }  . \label{theta}
\end{equation}
Thus the nature of the denaturation transition is determined by
the temperature dependence of the fugacity  $z^*(w)$. This
behavior can be classified into three distinct regimes depending
on the value of the exponent $c$. These regimes are most easily
understood through a graphical
solution of (\ref{pole}).

\subsubsection*{Case (i): $c \le 1$. No phase transition}
\begin{figure}
\includegraphics[width=8cm]{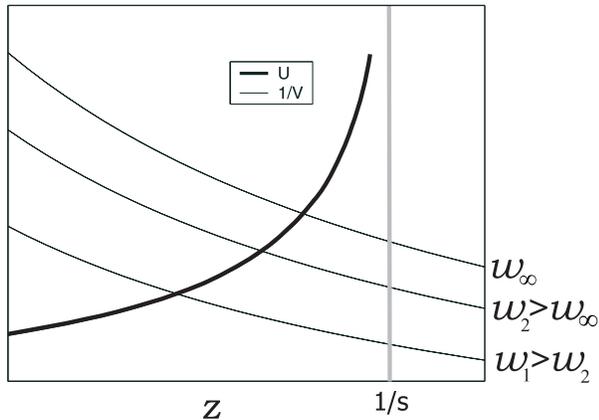} \caption{A typical behavior
of the functions $U$ and $1/V$ for $c<1$ (here $c=0.5$). See text
for explanation. \label{fig:c05}}
\end{figure}

A schematic representation of the graphical solution of
(\ref{pole}) in this case is given in Fig.~\ref{fig:c05}. The
function $U(z)$ is finite for any $z<z_{\rm M}=1/s$. Since the sum
\begin{equation}
U(1/s)=\sum_{\ell=1}^{\infty}\frac{1}{\ell^c}  , \label{sumcrit}
\end{equation}
diverges for $c \le 1$, the function $U(z)$ increases smoothly to
infinity as $z$ approaches $1/s$. For a given value of $z$ the
function $1/V(z)$ increases as the temperature increases. In
Fig.~\ref{fig:c05} $1/V(z)$ is plotted vs.~$z$ for three values of
$ w $. One can see that as the temperature is increased the
crossing point $z^*$ of the two graphs increases smoothly until it
saturates at $w= w _\infty$ ($T=\infty$). Therefore $\theta$
decreases smoothly as temperature is increased and no phase
transition takes place. In this case the strands are always bound
at all temperatures.

\subsubsection*{Case (ii): $1< c \le 2$. Continuous phase
transition} 
\begin{figure}
\includegraphics[width=8cm]{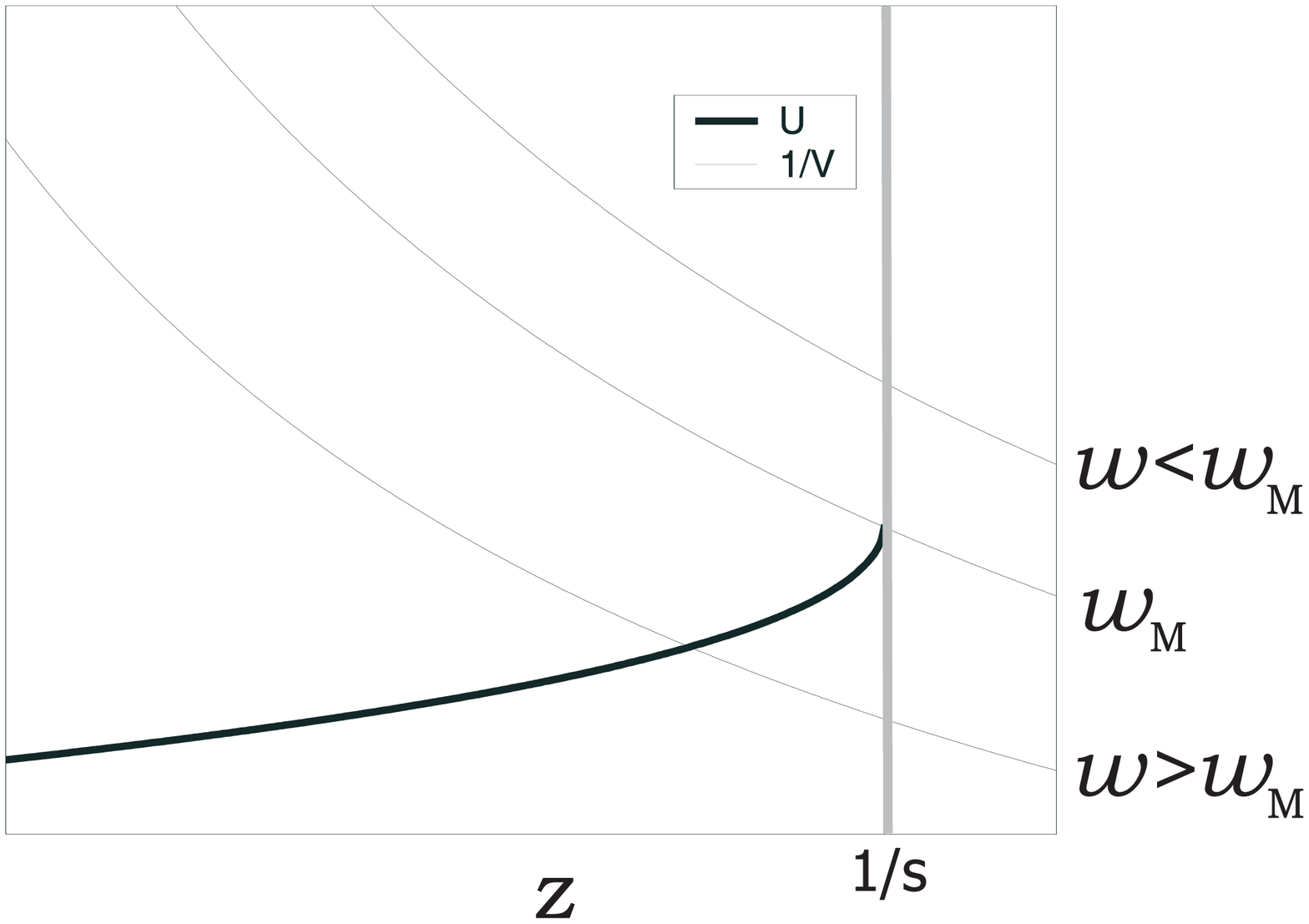} \caption{A typical behavior
of the functions $U$ and $1/V$ for $1<c \le 2$ (here $c=1.5$). See
text for explanation. \label{fig:c15}}
\end{figure}

A schematic representation of the graphical solution
of Eq.~(\ref{pole}) in this case is given in Fig.~\ref{fig:c15}.
Here the sum (\ref{sumcrit}) is finite at $z=z_{\rm M}=1/s$ since
$c>1$. The function $U(z)=\Phi_c(zs)$ increases smoothly to a
finite value as $z$ approaches $z_{\rm M}$ and becomes infinite
for $z>z_{\rm M}$. In Fig.~\ref{fig:c15} the function $1/V(z)$ is
plotted for three values of $ w $. One can see that as temperature
is increased $z^*$ increases until it reaches $1/s$ at $w = w_{\rm
M}$. Above the transition, for $w < w_{\rm M}$, $z^*$ remains
equal to $1/s$ in the thermodynamic limit. Here, the $\langle L
\rangle \to \infty$ limit is obtained through the divergence of
the factor $Q(z=1/s)$ in the numerator. A more careful analysis of
the denaturated regime is presented in
Appendix~\ref{hightemp:sec}. Note that for a transition to take
place one must have $1/V(z=1/s,w=1) \geq U(1/s)$. Otherwise there
is no phase transition and the two strands are bound at all
temperatures. Since $c \le 2$ the derivative of $U(z)$ diverges at
the transition. This implies that $\theta=\partial \ln z^*
/\partial \ln  w $ approaches zero continuously, yielding a
continuous transition. Since the derivative $\partial \ln z^*
/\partial \ln  w $ decreases with increasing $c$, the closer $c$
to two, the sharper the transition.

\subsubsection*{Case (iii): $c > 2$. First order phase
transition} 
\begin{figure}
\includegraphics[width=8cm]{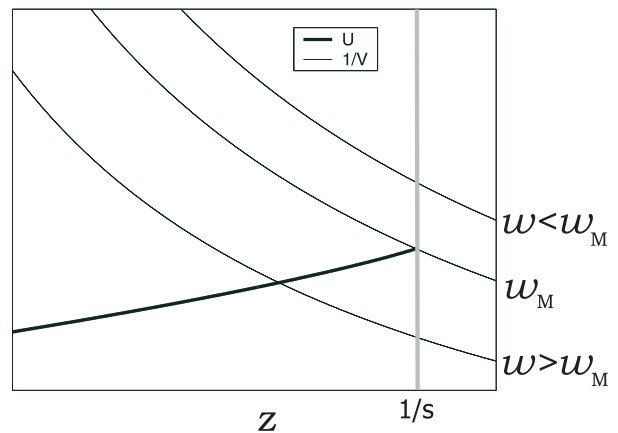} \caption{A typical behavior
of the functions $U$ and $1/V$ for $c>2$ (here $c=2.5$). See text
for explanation. \label{fig:c25}}
\end{figure}

A schematic representation of the graphical solution
of Eq.~(\ref{pole}) in this case is given in Fig.~\ref{fig:c25}.
Here both $U(z)$ and its derivative are finite at $z=z_{\rm
M}=1/s$. As in the previous case there is a transition for  $ w =
w_{\rm M}$. However, since the derivative of $U(z)$ is finite,
$\theta$ approaches a finite value as the transition is approached
from below. Above the transition $\theta$ vanishes in the
thermodynamic limit. The transition is therefore first-order. The
discussion of the high-temperature phase is again deferred to
Appendix~\ref{hightemp:sec}.

To summarize, three different scenarios exist depending
only on the value of $c$. These are
\begin{eqnarray*}
c \le 1{:} &\quad &\mbox{no phase transition;} \\
1<c \le 2{:} & \quad &\mbox{continuous phase transition;}\\
c>2{:} & \quad &\mbox{first-order phase transition.}
\end{eqnarray*}
The nature of the phase transition is thus directly related to the
number of configurations of long denaturated loops within the
chain. In the early studies of this problem the exponent $c$ was
evaluated by considering the number of random walks which return
to the origin as modeling the loop configurations \cite{PS2}. It
is easy to show that in $d$ dimensions the model yields  $c=d/2$.
This implies that there is no transition for $d \le 2$, a
continuous transition for $2<d \le 4$ and a first-order phase
transition for $d>4$. The model was subsequently extended to
include the repulsive short range interaction which exists between
the strands constituting a loop. In this approach the loop is
modeled as a self-avoiding walk~\cite{Fisher}. This yields $c=d
\nu$, where $\nu$ is the exponent associated with the radius of
gyration $R_{\rm G}$. For a self avoiding random walk of length
$L$ one has $R_{\rm G} \sim L^\nu$, with $\nu=3/4$ in $d=2$ and
$\nu \approx 0.588$ in $d=3$. This yields $c=3/2$ in $d=2$ and
$c=1.766$ in $d=3$. Thus  the transition is continuous in both
cases, although it is sharper than when the repulsive interaction
is neglected altogether.

The two estimates of the exponent $c$ described above treat the
loop as an isolated object and thus neglect its interaction with
the rest of the chain. This simplification seems essential, since the
formalism of the model relies on the segments composing the chain
as being independent. In the next section we show that the
repulsive interaction between a loop and the rest of the chain may
be accounted for. Although we treat these interactions only in an
approximate way, we are able to give insight into the unbinding
mechanism and on the nature of the transition.

\subsection{Excluded-volume effects}

\begin{figure}
\includegraphics[width=8cm]{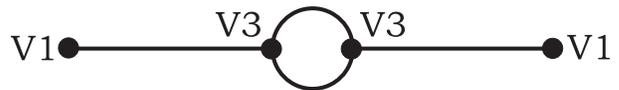} \caption{The topology of the loop embedded in a chain. The
length of the chain from a vertex of type $V1$ to the nearest
vertex of type $V3$ is $L$. The length of each of the two strands
connected to the $V3$ vertices is $\ell$. \label{fig:top1}}
\end{figure}

To account for the excluded volume interactions between a loop and
the rest of the chain we note that a microscopic configuration of
the DNA molecule is composed of many bound and unbound segments
of various length. When evaluating the number of available
configurations of a loop, one has to take into account the
interactions with all these bound and unbound segments. Here we
simplify the problem and neglect the internal structure of the
rest of the chain. We thus consider a loop embedded in a flexible
chain (see Fig.~\ref{fig:top1}) and study the number of
configurations of a chain endowed with this topology, assuming
that it is self avoiding. We will show that in the limit where the
loop length, $2\ell$, is much smaller than the length of the rest
of the chain, $2L$, the statistical weight of this topology can be
written as a product of the statistical weights of the loop and of
the chain. The weight of the loop is found to be of the same form
as that of a free loop but with a different exponent $c$. This
exponent is found to be larger than $2$ in dimensions $2$ and
above, yielding a first order denaturation transition.

To carry out this analysis we use results obtained by Duplantier
et al \cite{Dup,SFLD} for the number of configurations of polymer networks of
arbitrary topology. In order to make the paper self contained we
first review in some detail these results. They represent an
extension of the well known results for the number of
configurations of a simple self-avoiding random walk \cite{deGen}.
In this case it is known that the number of configurations scales
as
\begin{equation}
\Gamma_{\rm linear} \sim s^L L^{\gamma-1} , \label{onechain}
\end{equation}
where $L$ is the length of the polymer, $s$ is a non-universal
constant and $\gamma$ is a universal exponent. The exponent
$\gamma$ is known exactly in $d=2$, numerically in $d=3$ and via
an $\varepsilon$ expansion in $d=4-\varepsilon$. Above $d=4$
self-avoiding interactions becomes irrelevant and thus the number of
configurations of self-avoiding random walks scales as that of
ordinary random walks, yielding $\gamma=1$.
\begin{figure}
\includegraphics[width=8cm]{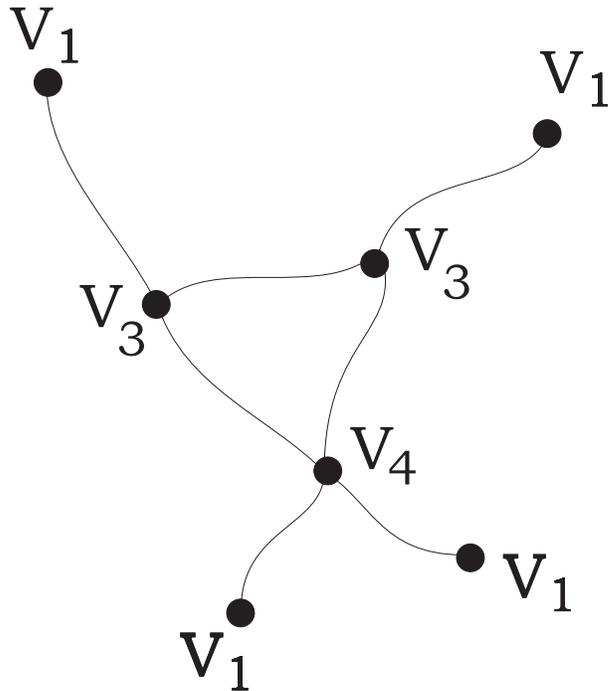} \caption{An example of a polymer network. The
network has four vertices $V_1$ of order 1, two vertices $V_3$ of
order 3 and one vertex $V_4$ of order 4. It also has one loop.
Thus in Eq. \ref{Dupexp} $n_1=4$, $n_3=2$, $n_4=1$ and ${\cal
L}=1$. \label{fig:network}}
\end{figure}

The generalization of this result to an arbitrary polymer network
goes as follows: Consider a branched self-avoiding polymer ${\cal
G}$ of arbitrary topology (see for example
Fig.~\ref{fig:network}). The polymer is made of $N$ chains of
lengths $\ell_1,\ell_2, \ldots \ell_N$. These are tied together at
vertices with different number of legs. A vertex with $k$ legs is
said to be of order $k$ ($k \geq 1$). The number of vertices of
order $k$ is denoted by $n_k$. The number of configurations of the
network, $\Gamma_{\cal G}$, is then given by
\begin{equation}
\Gamma_{\cal G} \sim s^L L^{\gamma_{\cal G}-1} g\left(
\frac{\ell_1}{L},\frac{\ell_2}{L}, \ldots,\frac{\ell_N}{L} \right)
, \label{Dupsca}
\end{equation}
where $L=\sum_i \ell_i$ is the total length of the network and $g$
is a scaling function. This expression is valid asymptotically as
$L\to \infty$, as long as all $\ell_i$'s diverge with the same
rate. We note that the relation is valid also when the persistence
length of each of the chains composing the graph is different.
This variance can be absorbed through a rescaling of the chains
lengths $\ell_i$. The exponent $\gamma_{\cal G}$ depends only on
the topology of the network and is given by
\begin{equation}
\gamma_{\cal G}= 1 - \nu d {\cal L} +\sum_{k \geq 1} n_k \sigma_k.
\label{Dupexp}
\end{equation}
Here ${\cal L}$ is the number of independent loops in the network,
$d$ the spatial dimension and $\nu$ is the exponent related to the
radius of gyration of a self-avoiding random walk. Each $k$-vertex
contributes its scaling dimension $\sigma_k$. Since vertices of
order $k$ appear $n_k$ times we obtain the term $n_k\sigma_k$.
Specifically, if all lengths $\ell_i$ scale in the same way, the
number of configurations scales as $\Gamma_{\cal G} \sim s^L
L^{\gamma_{\cal G}-1}$. The scaling dimensions $\sigma_k$, defined
for $k\ge 1$, are known exactly in $d=2$ from conformal
invariance:
\begin{equation}
\sigma_k=(2-k)(9k+2)/64,
\end{equation}
and to order $\varepsilon^2$ in $d=4-\varepsilon$:
\begin{eqnarray}
\sigma_k &=&(\varepsilon /8 )(2-k)k/2 \nonumber\\
&&{}+ (\varepsilon /8)^2 k(k-2)(8k-21)/8 + O(\varepsilon^3).
\end{eqnarray}
Also, good estimates for the values of the exponents in $d=3$ are
available through Pad\'{e} and Pad\'{e}-Borel approximants.
Clearly $\sigma_2=0$ as one would expect and above $d=4$, where
the self-avoiding interaction is irrelevant, all the exponents
$\sigma_k$ are zero.

Consider now the topology depicted in Fig.~\ref{fig:top1}. We are
interested in finding the number of configurations of the network
in the limit $\ell \ll L$, when the loop size is much smaller than
the length of the rest of the chain. Using the results by
Duplantier (see Eq.~(\ref{Dupsca})), the number of configurations
can be written as
\begin{equation}
\Gamma \sim s^{L+\ell} (L+\ell)^{\gamma_{\rm loop}-1} g(\ell/L),
\label{scaling}
\end{equation}
for large $L$ and $\ell$. Here $g(x)$ is a scaling function and
$\gamma_{\rm loop}$ can be evaluated using Eq.~(\ref{Dupexp}). For
the topology considered above of a loop embedded in two segments
(Fig.~\ref{fig:top1}) we have: one loop, ${\cal L}=1$; two vertices of order
one, $n_1=2$, corresponding to the two free ends of the chain
(denoted by $V1$ in the figure); and two vertices of order three,
$n_3=2$ (denoted in the figure by $V3$). Using Eq.~(\ref{Dupexp})
we obtain
\begin{equation}
\gamma_{\rm loop}=1- d \nu + 2 \sigma_1 + 2 \sigma_3.
\label{gloop}
\end{equation}
The limit of interest is that of a loop size much smaller than the
length of the chain, $\ell/L \ll 1$. Clearly, in the limit $\ell/L
\to 0$, the number of configurations should reduce to that of a
single self-avoiding open chain, which, to leading order in $L$,
is given by $s^{L} L^{\gamma-1}$, where $\gamma=1+ 2\sigma_1$.
This implies that in the limit $x \ll 1$
\begin{equation}
g(x) \sim x^{\gamma_{\rm loop}-\gamma}. \label{limit}
\end{equation}
Thus the $\ell$-dependence of $\Gamma$, which gives the change in
the number of configurations available to the loop, is given by
\begin{equation}
\Gamma \sim s^\ell \ell^{\gamma_{\rm loop}-\gamma} \cdot s^L
L^{\gamma-1}. \label{gdep}
\end{equation}
It is therefore evident that, for large $\ell$ and $L$ and in the
limit $\ell/L \ll 1$, the partition sum is decomposed into a
product of the partition sums of the loop and that of the rest of
the chain. The excluded volume interaction between the loop and
the rest of the chain is reflected in the value of the effective
exponent $c$. This result is very helpful since it enables one to
extend the Poland-Scheraga approach described in the previous
section to the case of interacting loops. From Eq.~(\ref{gdep})
one sees that the appropriate effective exponent $c$ is given by
\begin{equation}
c = \gamma-\gamma_{\rm loop}=d\nu-2\sigma_3 . \label{cdef}
\end{equation}
In $d=2$, $\sigma_3=-29/64$~\cite{Dup} and $\nu=3/4$, yielding
\begin{equation}
c= 2+13/32 \;. \label{2d}
\end{equation}
In $d=4-\varepsilon$ to $O(\varepsilon^2)$, one has
$\sigma_3=-3\varepsilon/16+9\varepsilon^2/512$ and
$\nu=1/2(1+\varepsilon/8+15/4(\varepsilon/8)^2)$, yielding
\begin{equation}
c= 2+\varepsilon/8+ 5 \varepsilon^2/256\;. \label{4d}
\end{equation}
In $d=3$, one may use Pad\'{e} and Pad\'{e}-Borel approximations to
obtain $\sigma_3 \approx -0.175$ \cite{SFLD} which with the value
$\nu \approx 0.588$~\cite{SFLD} yields
\begin{equation}
c \approx 2.115.
\end{equation}
The value of the exponent $c$ is unaffected by the different
persistence length of a bound and unbound DNA segment. This is
since as stated above Eq.~(\ref{Dupsca}) is valid
 also when the persistence length of different polymers composing
the network are different.
Equation~(\ref{cdef}) can be understood intuitively by remarking
that taking the limit $\ell/L \to 0$ corresponds to shrinking the
loop. By doing so one loop and two vertices of order $3$ are
eliminated. The exponent $c$ is the difference between the
exponent $\gamma_{\cal G}$ of the network after shrinking the loop
and the same exponent before the loop has shrunk. Therefore $c$
gets a contribution of $d \nu$ from the eliminated loop and $-2
\sigma_3$ from the two vertices of order $3$.

\begin{figure}
\includegraphics[width=8cm]{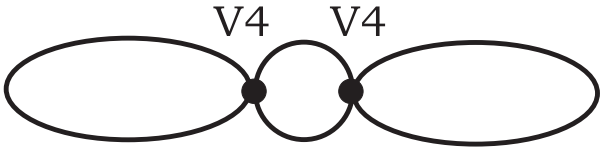} \caption{An extreme topology where the loop of length $2l$ is
embedded in two denaturated loops of size $2L$ each. The vertices
of order $4$ are denoted by $V4$. \label{fig:top2}}
\end{figure}

As stated above the rest of the chain is in fact composed of both
bound and unbound segments. This structure has been neglected in
the above analysis. To estimate the effect of the denaturated
segments we consider the extreme case in which the rest of the
chain is fully denaturated. That is, a loop embedded within two
large loops each of size $2L$ (see Fig.~\ref{fig:top2}). An
analysis similar to the one presented above yields for the value
of $c$,
\begin{eqnarray}
c &=& d\nu- \sigma_4 \;, \label{top2}\\
&=& 2 + 11/16\;,
\;\;\;\;\;\;\;\;\;\;\;\;\;\;\;\;\;\;\;\; {\rm in} \;\; d=2 \;,
\\
&=& 2+\varepsilon/4- 15 \varepsilon^2/128 \;, \;\;\;\;\; {\rm in}
\;\; d=4-\varepsilon,
\end{eqnarray}
where the values $\sigma_4=-19/16$ in $d=2$ and
$\sigma_4=-\varepsilon/2+11(\varepsilon^2/8)^2$ in
$d=4-\varepsilon$ dimensions~\cite{Dup} are used along with those
of $\nu$. Using $\sigma_4 \approx -0.46$ obtained by Pad\'{e} and
Pad\'{e}-Borel approximations gives in $d=3$ the value $c \approx
2.22$. Therefore, the effect of the extra excluded volume
interaction is to increase the value of $c$. It is easy to
check that the exponent governing the $\ell$-dependence
remains unchanged even if the end points of one or both
the outer loops are set to be unbound.

In both topologies considered above the value of the exponent $c$
is larger than $2$ in $d=2$, $d=4-\varepsilon$ and $d=3$ and
strongly suggests that the transition is {\it first order} for any
$d \ge 2$. Our analysis assumes that the size of the loops in the
system is much smaller than the total chain length. The fact that
the transition is first order implies that the loop size remains
finite. This makes the analysis self-consistent. Note that the
loop size distribution, $P(\ell)$, is rather broad at the
transition and behaves, for large $\ell$, as
\begin{equation}
P(\ell) \sim \frac{1}{\ell^c} \label{loopdist}.
\end{equation}
Thus high enough moments of the loop size distribution always
diverge. Although the transition is first order for $c>2$ it
exhibits some critical properties. Our analysis suggests that $2 <
c< 3$, so that already the variance of the loop size is predicted
to diverge. A recent numerical study of the loop size
distribution, where the excluded volume interactions have been
fully taken into account, has verified this prediction with a
measured value of $c=2.10 \pm 0.02$ \cite{COS}.

The exponent $c$ is an effective exponent calculated by assuming a
specific structure of the whole chain. In fact for any
configuration, as long as the bound segments connected to the
denaturated loop under consideration are long, the value of the
exponent $c$ remains the same as that given by (\ref{cdef}) (see
Fig.~\ref{fig:top1}). This is due to the fact that shrinking the
loop eliminates one loop and two vertices of order $3$. This
result is also easy to verify using a scaling argument similar to
the one presented above. The value of the exponent $c$ is altered
to (\ref{top2}) (see Fig.~\ref{fig:top2}) when one of the segments
connected to the loop is short. This may be verified by noticing
that as the loop is shrunk a vertex of order $4$ and one loop are
lost. Note that this does not imply that these values of $c$ can
be used to give bounds on the true partition function of the
model.

\subsection{The end segment distribution}\label{endseg:sec} In the
previous section it was argued that the excluded volume
interactions between a denaturated loop and the rest of the chain
cause the transition to be first order. As was noted this result
is unaffected by the boundary terms $V_0(z)$ and $Q(z)$ (see
Eq.~(\ref{UV})). However, when the DNA molecule is fully
denaturated, its statistical properties are determined by the
boundary term $Q(z)$. This is evident as the entropy of a
denaturated molecule with free ends is much higher than that of
one with the ends constraint to meet each other. Namely, the
number of configurations of a self-avoiding random walk is much
higher than that of a self-avoiding random walk which returns to
the origin. Thus, one expects the average length of the end
segment to diverge at the transition. This does not affect our
previous results as long as the thermodynamic limit, $L \to
\infty$, is taken before the temperature $T$ approaches the
melting temperature $T_{\rm M}$. In this case the size of the end
segment is always much smaller than the rest of the chain.

The average length of the end segment is given by
\begin{equation}
\xi = \left.z\frac{\partial \ln Q}{\partial z}\right|_{z=z^*}.
\end{equation}
where $z^*$ is evaluated using Eq.~(\ref{pole}). The behavior of
$\xi$ near the transition takes one of three forms depending on
the numerical value of $\bar{c}$:

\noindent ({\em i}) $\bar c \le 1$. In this case
$Q(sz)=1+\Phi_{\bar c}(sz)$ diverges like $|z-z_{\rm M}|^{\bar
c-1}$ (see Appendix \ref{polylog:sec}) as $z\to z_{\rm M}=1/s$.
Clearly its derivative with respect to $z$ diverges like
$|z-z_{\rm M}|^{\bar c-2}$ and hence the average length of the end
segment diverges like $\xi \sim |z^*-z_{\rm M}|^{-1}$.

\noindent ({\em ii}) $1<\bar{c} \le 2$. Here $Q(sz)$ is finite but
its derivative diverges like $|z-z_{\rm M}|^{\bar{c}-2}$ as $z\to
z_{\rm M}=1/s$. Thus, the average length of the end segment
diverges like $\xi \sim |z^*-z_{\rm M}|^{\bar{c}-2}$.

\noindent ({\em iii}) $\bar{c} > 2$. In this case both $Q(sz)$ and
its derivative are finite at the transition. Hence the end segment
length is finite at the transition.

We next use the temperature dependence of $z^*(w)$ in order to
evaluate the behavior of the end segment length $\xi$ as a
function of the temperature near the transition. Here one finds
two regimes depending on the value of the exponent $c$ related to
the entropy of the bulk loops \cite{Fisher}. To derive these
relations we note that for $z^*$ close to $z_{\rm M}$
Eq.(\ref{singularity}) yields for $c>1$,
\begin{equation}
U(z_{\rm M})-U(z^*) \sim |z_{\rm M}-z^*|^\zeta,
\end{equation}
where $\zeta=\min(1,c-1)$. Using $ w(z)=[z(1+U(z))]^{-1}$ from
Eq.~(\ref{transtion}) we obtain
\begin{equation}
U(z_{\rm M})-U(z^*)=\frac{1}{ w_{\rm M}z_{\rm M}}-\frac{1}{ w
(z^*) z^*},
\end{equation}
where $w_{\rm M}=w(z_{\rm M})$. Rewriting this expression yields
\begin{equation}
|z_{\rm M}-z^*|^\zeta  \sim  a w_{\rm M}(z^*-z_{\rm M}) + a ( w
(z^*)- w_{\rm M})z^* , \label{ztoome}
\end{equation}
where $a=1/(w_{\rm M} z_{\rm M} w (z^*) z^*)$ is finite at the
transition. Clearly $w(z^*)- w_{\rm M}  \sim   T_{\rm M} - T $
near the transition, where $T_{\rm M}$ is the melting temperature
and $T$ is the temperature corresponding to $ w (z^*)$. Thus,
Eq.~(\ref{ztoome}) yields two regimes for the temperature
dependence of $z^*$. For $1<c \le 2$ the linear term in $(z^* -
z_{\rm M})$ in Eq.(\ref{ztoome}) is negligibly small compared to
the left hand side and hence
\begin{equation}
z_{\rm M}-z^* \sim \vert T-T_{\rm M} \vert^{1/(c-1)}.
\end{equation}
On the other hand for $c >2$ $\zeta=1$ and therefore
\begin{equation}
z_{\rm M}-z^* \sim \vert T-T_{\rm M} \vert. \label{zT}
\end{equation}

Thus, the behavior of the end segment length $\xi$ depends on both
exponents $c$ and $\bar{c}$. We have already shown that $c>2$. We
now turn to estimate the exponent $\bar{c}$. This exponent is
associated with the degeneracy , $s^\ell / \ell^{\bar{c}}$, of an
end segment of length $\ell$, and its value may be deduced using
the scaling argument presented above. One considers a Y-fork
topology where the two denaturated strands are much smaller than
the bound segment representing the rest of the chain. Using the
terminology introduced in the preceding subsection  one has to
consider the difference of the exponent corresponding to the
Y-fork topology and the one corresponding to linear chain. This
yields
\begin{equation}
\bar{c}=-(\sigma_1 + \sigma_3). \label{cbar}
\end{equation}
In $d=2$, $\sigma_1=11/64$ and $\sigma_3=-29/64$, so that
$\bar{c}=9/32$. In $d=3$, one may use the Pad\'{e} and
Pad\'{e}-Borel approximation results $\sigma_1 \simeq 0.083$ and
$\sigma_3 \simeq -0.175$ to obtain $\bar{c} \simeq 0.258$. Finally
in $d=4-\varepsilon$, $\bar{c}=\varepsilon /8 + O(\varepsilon^2)$
while above four dimensions clearly $\bar{c}=0$. This suggests
that the exponent $\bar{c}$ is smaller than one for any $d \ge 2$.

Since the degeneracy exponents satisfy $c>2$ and $\bar{c}<1$ the
analysis presented above suggests that near the melting transition
the end segment length diverges like
\begin{equation}
\xi \sim \frac{1}{\vert T-T_{\rm M} \vert} .
\end{equation}

\subsection{Summary and overview of thermal denaturation }

In the preceding subsections we reviewed the analysis of Poland
and Scheraga, showing that the nature of the denaturation
transition is governed by the exponent $c$ associated with the
entropy of a denaturated loop embedded in a  DNA molecule. It is
found that for $c \le 1$ there is no transition, for $1<c \le 2$
there is a continuous transition, while for $c>2$ the transition
is first order. Using recent results for the number of
configurations of a self avoiding  polymer network of general
topology we have demonstrated that $c>2$ in $d=2$ dimensions and
above. While the treatment is approximate the results strongly
suggest that the transition is indeed first order in these
dimensions. The results are unchanged even if the different
persistence length of the bound and denaturated configurations is
taken into account.

The analysis of the length distribution of the end segment shows
that the molecule melts from the unbound end. The boundary
condition used in this analysis is such that the two strands are
bound at one end. It is easy to show that when the boundary
condition is modified such that the two strands are bound at the
center rather than at one end, melting takes place from both ends.

It is interesting to consider the case of a homopolymer  where all
the bases on one strand are the same. For example a molecule in
which one strand is made only of G bases and the other strand is
made only of C bases. Here each base, and therefore a monomer in
the model, can bind to any other base on the other strands and not
just to a specific monomer, as is the case for a heteropolymer.
The two arms of a loop need not be of the same length. Thus the
number of configurations of a loop of length $\ell$ changes by a
factor of $\ell$. This amounts to using $\ell \cdot s^\ell
/\ell^{c}$ for the weight of a loop, effectively reducing the
exponent $c$ to $c-1$. Since we have shown that $2<c<3$ for $d \ge
2$, this implies for a homopolymer the effective $c$ is smaller
than $2$ but greater than $1$. The transition in this case is thus
expected to be {\em continuous}.

Finally, we comment on recent attempts to account for the
first-order nature of the denaturation transition without having
to resort to excluded volume interactions. Within this approach
the two strands are considered as directed polymers and thus they
do not self intersect. Let $V(\vec r)$ the interaction potential
between the corresponding monomers on the two strands. It is
repulsive at short distances, has an attractive well at the
characteristic pair bond distance and it tends to zero at large
distances. Using a transfer matrix approach, the thermodynamic
properties of the chain are obtained by the ground state
$\psi_0(\vec r)$ wavefunction of a Schr\"{o}dinger equation which
takes the form \cite{deGen}
\begin{equation}
-\frac{1}{2 m(\vec r)} \nabla^2 \psi_0(\vec r)+ V(\vec
r)\psi_0(\vec r) =\epsilon_0 \psi_0(\vec r). \label{Schr}
\end{equation}

Here $\epsilon_0$ is the ground state eigenvalue associated with
the wave function $\psi_0(\vec r)$. The mass $m$ represents the
stiffness of the chain. The $\vec r$-dependence of the mass is
taken to account for the change of stiffness between bound and
unbound strands~\cite{TDP}. One usually assumes that the potential
$V(\vec r)$ is short ranged (exponentially decaying with $\vec r$
at large distances) and that $m(\vec r)$ varies with $\vec r$ over
the same distance. The probability density of finding the strands
a distance $\vec r$ from each other is given by $\vert \psi_0(\vec
r) \vert^2$. In this language the order parameter $\theta$ is
given by
\begin{equation}
\theta=\frac{\int_0^a \vert \psi_0(r) \vert^2 d^d r
}{\int_0^\infty \vert \psi_0(r) \vert^2 d^d r}, \label{Schtheta}
\end{equation}
where $a$ is the range of the binding potential $V(\vec r)$.
Here, for simplicity, we assume that neither $m(\vec r)$
nor $V(\vec r)$ depend on the orientation.

The denaturation transition occurs when $\epsilon_0$ reaches zero
so that there is no longer a bound state in the system. A first
order transition occurs if $\theta$ has a non-zero value at that
point. In this case one should have a bound state with energy
$\epsilon_0=0$. Note that although $\theta$ might jump at the
transition, the average distance between the strands, which is not
the order parameter of the system, might diverge continuously.
Thus, we are interested in the solution of the equation
\begin{equation}
-\nabla^2 \psi_0(\vec r)+ 2 m(r) V(r)\psi_0(\vec r)=0.
\label{Sch}
\end{equation}
Since $V(r)$ decays exponentially with $r$, the asymptotic
behavior of $\psi_0(\vec r)$ at large distances is obtained by the
equation
\begin{equation}
\nabla^2 \psi_0(\vec r)=0. \label{Schlarge}
\end{equation}
It is easy to show that this implies $\psi_0(\vec r) \sim
1/r^{d-2}$ for large $r$. That is, the wave function for large $r$
is normalizable only for $d>4$ so one has a localized zero energy
ground state. Therefore using Eq.~(\ref{Schtheta}) one can see
that for any dimension $d<4$ the order parameter $\theta$ vanishes
at the transition as is expected for a continuous transition. A
first-order phase transition occurs only for $d>4$. Note that this
result is equivalent to using the Poland-Scheraga model with
random walks modeling the loop entropy. Within this approach the
transition may thus be altered to first order in lower dimensions
{\it only} when $V(r)$ are long-range, in contrast to recent
claims \cite{TDP}.

Recently, it has been argued that taking into account
excluded-volume interactions {\it between} the two strands while
neglecting these interactions within each strand effectively leads
to a long range potential between the two strands \cite{GMO}. This
model, which takes into account the excluded volume interactions
only partially also leads to a first order transition.

\section{The unzipping transition}
The introduction of new and powerful techniques, such as optical
tweezers~\cite{tweezers} and atomic force microscopes~\cite{AFM},
has made possible the manipulation of single biological
macromolecules. A number of experiments have investigated the
response of double-stranded DNA to external forces and
torques~\cite{DNA}. Recently, it has become possible to apply and
measure a force pulling apart two strands of a DNA double
helix~\cite{Brockelmann}. Previous theoretical studies of the
unzipping transition have been carried out using the directed
polymer approach where self-avoiding interactions are not
accounted for~\cite{BHA,LN,LN2,MIM,CMM}. In most of these studies
the unzipping of homopolymers has been analyzed. Heterogeneous
chains have also been considered in some of these
studies~\cite{LN}. In this section we extend the analysis of the
PS model to consider the unzipping of homopolymers with
self-avoiding interactions. We show that the unzipping transition
is first order. We also calculate the dependence of the critical
unzipping force on the temperature at low forces, namely near the
melting temperature.

We consider a configuration where the corresponding  monomers at
one end of the chain are bound together, while a force $\vec f$ is
applied on the two monomers at the other end of the chain, pulling
the two strands apart. In this setup, the grand canonical
partition function takes the form
\begin{equation}
{\cal Z}=\frac{V_0(z)O(z)}{1-U(z)V(z)},
\end{equation}
where the factor $O(z)$ is the grand partition function of the
open tail under force. We have
\begin{equation}
O(z)=1+\sum_{\ell=1}^\infty {\cal Z}_{\rm end}(\ell) z^\ell,
\end{equation}
where ${\cal Z}_{\rm end}(\ell)$ is the canonical partition
function of an open end composed of two strands, each of length
$\ell$.

To evaluate ${\cal Z}_{\rm end}(\ell)$ we note that when no force
is applied the partition sum takes the form
\begin{equation}
{\cal Z}_{\rm end}(\ell) = \Lambda(2\ell) \sim \frac{s^\ell
}{\ell^{\bar c} },
\end{equation}
where, as discussed in sec.~\ref{endseg:sec}, $\bar c$ is given by
Eq.~(\ref{cbar}). When a force $\vec f$ is applied, we have
\begin{equation}
{\cal Z}_{\rm end}(\ell) =  \Lambda(2\ell) \int d\vec
r\,p_\ell(r)\,\exp(\vec f \cdot \vec r/T),
\end{equation}
where $p_\ell(r)$ is the probability distribution of the
end-to-end distance in the absence of a force. Turning to angular
coordinates we obtain
\begin{equation}
{\cal Z}_{\rm end}(\ell) =\Lambda(2\ell) \,{\cal I}_\ell(f/T),
\end{equation}
where
\begin{eqnarray}\label{intzeta:eq}
{\cal I}_\ell(f/T)  & = & S \int_0^\pi d\phi\, \sin^{d-2}\phi
\int_0^\infty
dr\,r^{d-1}\nonumber\\
&& {} \times p_\ell(r)\,\exp(fr\cos\phi/T),
\end{eqnarray}
in which $S$ is a constant which depends on dimensionality. We
assume that $p_\ell(r)$ has the same scaling form as that of
linear polymers
\begin{equation}
p_\ell(r)=R^{-d} \hat p(r/R).
\end{equation}
Here $R$ is a scaling length related to $\ell$ by
\begin{equation}
R \simeq R_0\, \ell^{\nu},
\end{equation}
where $\nu$ is the correlation length exponent of a linear
polymer.

We are interested in the behavior of $\hat p(x)$ at $x \gg 1$ (see
below). We assume that in this limit $\hat p(x)$ takes a form
similar to that corresponding to a linear polymer~\cite{desclo}
\begin{equation}
\hat p(x) = P \,x^\mu \,\exp(-D x^{\lambda}).
\end{equation}
Here $P$ and $D$ are constants, $\lambda=1/(1-\nu)$ and the
exponent $\mu$ is given by
\begin{equation}
\mu= (d/2+\nu d - \bar c)/(1-\nu). \label{kappa}
\end{equation}
This result can be obtained by applying the same reasoning used to
derive the corresponding expression for linear polymers to the
Y-fork configurations which are of interest here
\cite{desclo,moore}. Substituting into expression
(\ref{intzeta:eq}) we obtain
\begin{eqnarray}
\label{Il}
{\cal I}_\ell(f/T) & \propto & \int_0^\pi
d\phi\,\sin^{d-2}\phi\int_0^\infty dx\,
x^{d-1+\mu}\nonumber\\
&& {}\times \exp\left(-D x^{1/\lambda}+u x \cos\phi\right),
\end{eqnarray}
where $u=fR/T$. This expression is valid provided the integral is
dominated by large values of $x$, which is the case for $u \gg 1$.
This appears to be the relevant regime for piconewton forces. In
this situation we can evaluate the integral by steepest descent
and obtain the saddle-point equations
\begin{eqnarray}
\phi^* &=& 0;\\
x^* &=& \left(\frac{\lambda u}{D}\right)^{1/(\lambda-1)}.
\end{eqnarray}
They correspond to the non-Hookean elongation regime, described by
de~Gennes~\cite[p.~47ff]{deGen}. Note, that $x^*$ scales as
$u^{1/(\lambda-1)}$ and that $\lambda > 1$. Since we are
interested in the limit of $u \gg 1$ this justifies our choice of
using the tail of the distribution $\hat p(x)$. Therefore
\begin{equation}
{\cal I}_\ell(f/T)  \propto \ell^{\mu
(1-\nu)+\frac{d}{2}(1-2\nu)}\, \exp\left(A (f
R_0/T)^{1/\nu}\,\ell\right),
\end{equation}
where $A$ is a constant. Using this expression we have
\begin{equation}
O(z) \simeq 1+\sum_{\ell=1}^\infty \frac{%
\left[z s \exp\left(A (f R_0/T)^{1/\nu}\right)\right]^\ell}{%
\ell^{\bar \mu}}, \label{endpar}
\end{equation}
where the exponent $\bar \mu$ is given by
\begin{equation}
\bar\mu=-\bar c-\mu (1-\nu)-\frac{d}{2}(1-2\nu).
\end{equation}
Substituting $\mu$ from Eq.~(\ref{kappa}) into this expression
we obtain $\bar\mu=0$.

According to Eq.~(\ref{endpar}) at temperatures below the melting
temperature $T_{\rm M}$ the end segment partition sum $O(z)$
diverges at a critical, unzipping force $f_{\rm U}$, given by
\begin{equation}\label{critical:eq}
e^{-A (f_{\rm U} R_0/T)^{1/\nu}}=s z^*(w).
\end{equation}
Here $z^*(w)$ is the solution of Eq.~(\ref{pole}) (corresponding
to an infinitely long polymer). At this point the average length
of a loop in the bulk is finite. Hence the unzipping transition is
first order.

Near the transition, the length $\xi$ of the end segment diverges
like $|z^*-z_{\rm U}|^{-1}$, where $z_{\rm U}=e^{-\kappa (f_{\rm
U}/T)^{1/\nu}}/s$. This is a result of the fact that the exponent
$\bar \mu$ is smaller than $1$ (in fact it vanishes, as we have
seen). Since $z^*$ is regular in $f$, we have $\xi \sim |f-f_{\rm
U}|^{-1}$ or $\xi \sim |T-T_{\rm U}(f)|^{-1}$. Thus the two
strands separate gradually from the end as the critical force is
approached. Nonetheless the unzipping transition is first order.
The reason is that the transition takes place at a temperature
below the denaturation melting temperature $T_M$ where the loop
size distribution in the interior of the chain decays
exponentially with the loop size. Thus at this point the average
loop size in the interior of the chain is finite. On the other
hand the length of the end segment is finite as long as $f <
f_{\rm U}$ and its contribution to the order parameter $\theta$
and to the entropy is negligible. Therefore both the order
parameter and the entropy exhibit a discontinuity  to their values
in the unzipped state at the transition. Let us remark that
Eq.~(\ref{critical:eq}) implies that the critical force $f_{\rm
U}$ behaves like
\begin{equation}
f_{\rm U} \sim |T-T_{\rm M}|^\nu
\end{equation}
as $T\to T_{\rm M}$, at least as long as the forces are not too
small (so that the $u \gg 1$ limit in (\ref{Il}) is valid).

\section{Summary}
In this paper we have extended the Poland-Scheraga type models
introduced in the late fifties to fully take into account the
effect of self-avoiding interactions on the DNA denaturation
transition. We have shown that the model yields a first-order
transition in agreement with experiments and with numerical
simulations. Although the transition is thermodynamically first
order it exhibits critical behavior in some of its properties,
such as its loop size distribution and the length of the end
segment. It would be of interest to study these properties
experimentally to test these predictions.

We have also studied the unzipping transition of DNA and found it to be
first order. We have evaluated the behavior of the unzipping force near the melting point.

\begin{acknowledgments}
We are indebted to M. J. E. Richardson and H. Orland for many
inspiring discussions and their involvement in the early stages of
our study of the DNA denaturation transition. We also thank
discussion and suggestions by B. Duplantier. The work of LP was
partially supported by a Michael Visiting Professorship of the
Weizmann Institute and has been performed within a joint
cooperation agreement between Japan Science and Technology
Corporation (JST) and Universit\`a di Napoli \lq\lq Federico
II\rq\rq.
\end{acknowledgments}

\appendix
\section{Properties of the polylog function}\label{polylog:sec}
We summarize here a few elementary properties of the polylog
function that are used in the text~\cite[Sec.~1.11,
p.~27ff.]{Erdely}. The polylog function $\Phi_c(z)$ is defined by
the series
\begin{equation}\label{polylog:series}
\Phi_c(z)=\sum_{\ell=1}^\infty \frac{z^\ell}{\ell^c},
\end{equation}
which converges for $|z|<1$. For $|z|<1$ and $\Re {\rm e} \; c >0$
it has the integral representation
\begin{equation}\label{polylog:integral}
\Phi_c(z)=\frac{1}{\Gamma(c)}\int_0^\infty
dt\,t^{c-1}\,\frac{ze^{-t}}{1-ze^{-t}},
\end{equation}
where $\Gamma(c)$ is Euler's gamma function. From
Eq.~(\ref{polylog:integral}) it is easy to see that $\Phi_c(z)$
diverges like $|z-1|^{c-1}$ for $z\to 1$, if $c\le 1$, and that,
if $c>1$ and $1-z=\epsilon \ll 1$, one has
\begin{equation}\label{singularity}
\Phi_c(1)-\Phi_c(1-\epsilon) \sim \epsilon^\zeta,
\end{equation}
where the exponent $\zeta$ is equal to $\min(1,c-1)$. From the
series definition (\ref{polylog:series}) it is also evident that
\begin{equation}\label{derivative}
z\frac{d\Phi_c(z)}{dz}=\Phi_{c-1}(z).
\end{equation}

\section{High-temperature phase}\label{hightemp:sec}
Above the transition, i.e., for $w < w_{\rm M}$, the fugacity $z$
becomes $z_{\rm M}=1/s$ in the thermodynamic limit. In this
Appendix we discuss in some detail how the thermodynamic limit is
taken.

The value of the fugacity for a chain of length $L$ is obtained by
solving Eq.~(\ref{fugacity:eq}) for the fugacity $z$. It reads
\begin{equation}\label{length:eq}
\langle L \rangle = z\frac{V'_0(z)}{V_0(z)} +
z\frac{Q'(z)}{Q(z)}+z\frac{U'(z)V(z)+U(z)V'(z)}{1-U(z)V(z)}.
\label{Alength}
\end{equation}
For $w <  w_{\rm M}$, as $z$ approaches $z_{\rm M}$, the first
term is regular, while the second diverges as $|z-z_{\rm
M}|^{-1}$. In the third term the denominator does not vanish. The
term may or may not diverge depending on whether $U'(z)$ diverges
at $z_M$, namely, according to whether $c$ is smaller or larger
than 2. However, in any case this term is much smaller than the
second. We therefore have
\begin{equation}
|z-z_{\rm M}| \sim L^{-1}.
\end{equation}
Since the second term in Eq.~(\ref{length:eq}) is equal to the
number of units in the end segment, and since this is the
dominant term we conclude that almost all units belong to it in
the thermodynamic limit.

If both ends of the DNA chain are constrained to be bound, the
partition function assumes the form
\begin{equation}
{\cal Z}=\frac{V_0^2(z)}{1-U(z)V(z)}.
\end{equation}
Therefore the most singular term in Eq.~(\ref{length:eq}) is
absent. If $c\le 2$, $U'(z)$ diverges as $|z-z_{\rm M}|^{c-2}$, so
that the equation corresponding to (\ref{length:eq}) can still be
solved for any finite $L$. This means that the typical size of the
denaturated loops increases with increasing $L$.

In the case $c>2$, on the other hand, the derivative of $\ln {\cal
Z}$ with respect to $\ln z$ remains finite even for $z=z_{\rm M}$.
The chain thus seems to have a maximum length, $L_0(w)$,
corresponding to $z=z_{\rm M}$. Therefore, for longer chains one
has to consider the total length constraint more explicitly. To do
that we introduce a maximal length cutoff, $\bar{L}$, in the sums
(\ref{Udef}) and (\ref{Vdef}) defining $U(z)$ and $V(z)$,
respectively. With this cutoff the number of terms in each of
these series is finite, and thus they may be evaluated even for
$zs>1$. In this case the last term in Eq.~(\ref{Alength}) diverges
when $U(z)V(z)-1$ vanishes. Therefore, for any $\langle L\rangle$,
and particularly for $\langle L\rangle > L_0(w)$, one can find a
fugacity which satisfies $1-U(z)V(z) \sim 1/\langle L \rangle$.
Since for large $\bar{L}$, $U(z)$ grows exponentially as
$(sz)^{\bar{L}}$ this implies, to leading order in $\bar{L}$,
\begin{equation}
(sz)^{\bar{L}} \sim \left( 1 - \frac{a}{\langle L \rangle} \right)
\frac{1}{V},
\end{equation}
where $a$ is a constant. For large $\bar{L}$ the right hand side
of this equation approaches a constant, independent of $\bar{L}$.
This is due to the fact that for $z$ larger but sufficiently close
to $z_{\rm M}$ and at temperatures above the melting temperature
$V(z)$ is finite even for $\bar{L}$ going to infinity. In the
limit $\bar{L} \to \infty$ the fugacity behaves as
\begin{equation}
z=\frac{1}{s}+O\left( \frac{1}{\bar{L}} \right).
\end{equation}
Thus, in this case, $z$ approaches its limiting value from above.

\end{document}